\documentclass[aps,prd,10pt,twocolumn,superscriptaddress,showpacs,longbibliography,nofootinbib]{revtex4-1}

\pdfoutput=1
\usepackage{graphicx}
\usepackage{amsfonts,amsmath,amssymb,bm,bbm}
\usepackage{color}
\usepackage{enumitem}
\usepackage{url}  

\usepackage{hyperref}
\usepackage[capitalize]{cleveref}
\usepackage[dvipsnames, usenames]{xcolor}
\usepackage{orcidlink}

\usepackage{graphicx}
\newcommand{\orcid}[1]{\href{https://orcid.org/#1}{\includegraphics[width=10pt]{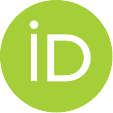}}}
\newcommand{\github}[1]{\href{https://github.com/#1}{\includegraphics[width=10pt]{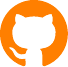}}}

\definecolor{romared}{RGB}{142,0,28}
\usepackage{hyperref}
\hypersetup{
    pdfnewwindow=true,      
    colorlinks=true,       
    linkcolor=romared,          
    citecolor=romared,        
    filecolor=romared,      
    urlcolor=romared        
}

\def\bi{\begin{itemize}[noitemsep,leftmargin=*]
\setlength\itemsep{1em}
        }
\def\ei{\end{itemize}}

\usepackage{lipsum}

\begin{document}

\title{High-energy neutrinos from choked-jet supernovae: Searches and implications}

\author{Po-Wen Chang \orcid{0000-0003-1134-0652}}
\email{chang.1750@osu.edu}
\affiliation{Center for Cosmology and AstroParticle Physics (CCAPP), Ohio State University, Columbus, Ohio 43210, USA}
\affiliation{Department of Physics, Ohio State University, Columbus, Ohio 43210, USA}

\author{Bei Zhou \orcid{0000-0003-1600-8835}}
\email{beizhou@fnal.gov}
\affiliation{William H. Miller III Department of Physics and Astronomy, Johns Hopkins University, Baltimore, Maryland 21218, USA}

\author{Kohta Murase \orcid{0000-0002-5358-5642}}
\email{murase@psu.edu}
\affiliation{Department of Physics, The Pennsylvania State University, University Park, Pennsylvania 16802, USA}
\affiliation{Department of Astronomy and Astrophysics, The Pennsylvania State University, University Park, Pennsylvania 16802, USA}
\affiliation{Center for Multimessenger Astrophysics, Institute for Gravitation and the Cosmos, The Pennsylvania State University, University Park, Pennsylvania 16802, USA}
\affiliation{School of Natural Sciences, Institute for Advanced Study, Princeton, New Jersey 08540, USA}
\affiliation{Center for Gravitational Physics and Quantum Information, Yukawa Institute for Theoretical Physics, Kyoto, Kyoto 606-8502 Japan}

\author{Marc Kamionkowski \orcid{0000-0001-7018-2055}}
\email{kamion@jhu.edu}
\affiliation{William H. Miller III Department of Physics and Astronomy, Johns Hopkins University, Baltimore, Maryland 21218, USA}

\date{\today}

\begin{abstract}
\noindent 
The origin of the high-energy astrophysical neutrinos discovered by IceCube remains largely unknown. Multimessenger studies have indicated that the majority of these neutrinos come from gamma-ray-dark sources. 
Choked-jet supernovae (cjSNe), which are supernovae powered by relativistic jets stalled in stellar materials, may lead to neutrino emission via photohadronic interactions while the coproduced gamma rays are absorbed. In this paper, we perform an unbinned maximum-likelihood analysis to search for correlations between IceCube's ten-year muon-track events and our SN Ib/c sample, collected from publicly available catalogs. In addition to the conventional power-law models, we also consider the impacts of more realistic neutrino emission models for the first time, and we study the effects of the jet beaming factor in the analyses. Our results show no significant correlation. Even so, the conservative upper limits we set to the contribution of cjSNe to the diffuse astrophysical neutrino flux still allow SNe Ib/c to be the dominant source of astrophysical neutrinos observed by IceCube. We discuss implications to the cjSNe scenario from our results and the power of future neutrino and supernova observations.
\end{abstract}

\maketitle

\section{Introduction}
\label{sec_introduction}

The discovery of high-energy (HE) astrophysical neutrinos by the IceCube Neutrino Observatory~\cite{IceCube:2013cdw,IceCube:2013low} has opened a new era of neutrino physics, astrophysics, and multimessenger astronomy. These neutrinos are nearly isotropic on the sky and have energies from 10~TeV to above the PeV scale, suggesting their sources to be the extragalactic populations of extreme cosmic accelerators. 
Searching for HE neutrino sources is also crucial to identifying sources of their parent HE cosmic rays and it offers unique opportunities to understand the acceleration mechanisms of the sources. 

\begin{figure}[t]
    \centering
    \includegraphics[width=0.75\columnwidth]{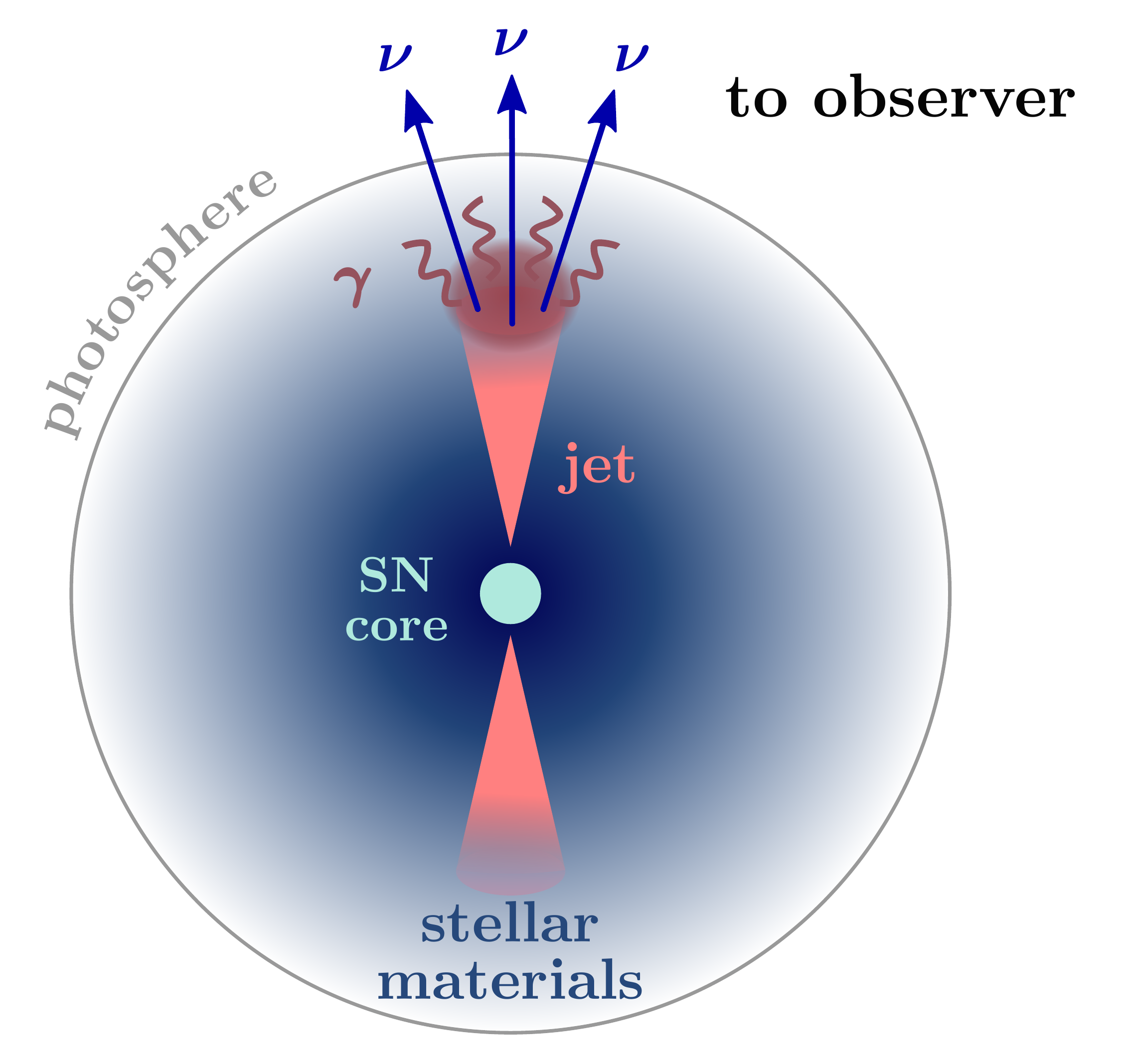}
    \caption{Schematic view of a core-collapse supernova with its jet choked inside the dense envelope of a progenitor star or external circumstellar materials. {\it While neutrinos and gamma rays are both produced by the cosmic rays inside the jet, only neutrinos can freely escape from the optically thick medium.}}
    \label{fig:cartoon}
\end{figure}

Various astrophysical objects have been studied as the sources of the HE neutrinos, such as gamma-ray bursts (GRBs)~\cite{Paczynski:1994uv, Waxman:1997ti, Dermer:2003zv, Kimura:2022zyg}, active galactic nuclei (AGN)~\cite{Berezinsky1977, Eichler:1979yy,PhysRevLett.66.2697,Murase:2015ndr}, supernovae (SNe)~\cite{Murase:2010cu,Murase:2017pfe}, tidal disruption events (TDEs)~\cite{Murase:2008zzc,Wang:2015mmh,Dai:2016gtz,Senno:2016bso,Lunardini:2016xwi}, and so on. 
Motivated by multimessenger observations of electromagnetic and gravitational-wave signals, significant efforts have been put into searching for the sources over the past decade. 
Compelling evidence for a few point sources has been reported: blazar TXS 0506+056~\cite{IceCube:2018dnn,IceCube:2018cha}, Seyfert II galaxy NGC 1068~\cite{IceCube:2019cia}, and the TDE candidates AT2019dsg, AT2019fdr, and AT2019aalc~\cite{Stein:2020xhk,Reusch:2021ztx,vanVelzen:2021zsm}. 
However, stacking analyses show that {\it none} of the aforementioned types of sources contribute a major fraction of the all-sky neutrino flux (a possible exception might be the nonjetted AGN if the observed 2.6$\sigma$ significance is interpreted as a signal~\cite{IceCube:2021pgw}). Furthermore, the measurement of the extragalactic gamma-ray background by the {\it Fermi} Large Area Telescope (LAT)~\cite{Fermi-LAT:2014ryh} has set robust constraints on gamma-ray-bright sources as dominant HE neutrino emitters, as a comparable flux of gamma rays are expected to be coproduced with neutrinos following $pp$ and $p\gamma$ interactions of cosmic rays~\cite{Murase:2013rfa}. 
Thus, it is likely that there is a class of HE neutrino sources opaque to GeV--TeV gamma rays, in which only neutrinos can freely escape from the cosmic-ray accelerators \cite{Murase:2015xka,Capanema:2020rjj,Capanema:2020oet,Fang:2022trf,Fasano:2021bwq}.

Choked-jet supernovae (cjSNe) are promising as such hidden neutrino sources~\cite{Meszaros:2001ms,Razzaque:2004yv,Razzaque:2005bh,Ando:2005xi,Murase:2013ffa,He:2018lwb}. \Cref{fig:cartoon} shows an illustration: in this scenario, the jets are stalled or ``choked'' inside the progenitor envelopes or circumstellar materials as they are not powerful enough. Gamma rays produced from cosmic rays accelerated in the choked jet are attenuated through optically thick environments below the stellar photosphere, leaving HE neutrinos as primary signals. Previous theoretical studies have shown that cjSNe could even explain all the HE astrophysical neutrinos observed by IceCube~\cite{Murase:2013ffa, Tamborra:2015fzv, Senno:2015tsn,Guetta:2019wpb,Carpio:2020app}.

In addition, cjSNe provide a unified scenario for Type Ib/c supernovae (SNe Ib/c), hypernovae, and the different types of GRBs, in which jet properties and shock-breakout conditions are crucial in making the difference~\cite{Nakar:2015tma,Senno:2015tsn}. 
The link to low-power (LP) GRBs such as low-luminosity (LL) GRBs and ultralong (UL) GRBs further strengthens the role of cjSNe as gamma-ray dark factories of HE neutrinos~\cite{Murase:2013ffa, Senno:2015tsn, Carpio:2020app}. From the observational side, many LP GRBs are missed by current GRB surveys, so current stringent limits on HE neutrino emission from classical, high-luminosity GRBs do not apply. 
From the theoretical side, the intrinsically weak jets associated with LP GRBs are more ideal for neutrino production than classical GRBs, as powerful jets generally lead to inefficient cosmic-ray acceleration in radiation-mediated shocks~\cite{Murase:2013ffa}.
Therefore, HE neutrinos also provide us with an essential tool to study the cjSN scenario and the observed GRB-SN connection~\cite{1998Natur.395..670G,Hjorth:2003jt,Stanek:2003tw,Modjaz:2015cca}.

\begin{figure}[t]
    \centering
    \includegraphics[width=\columnwidth]{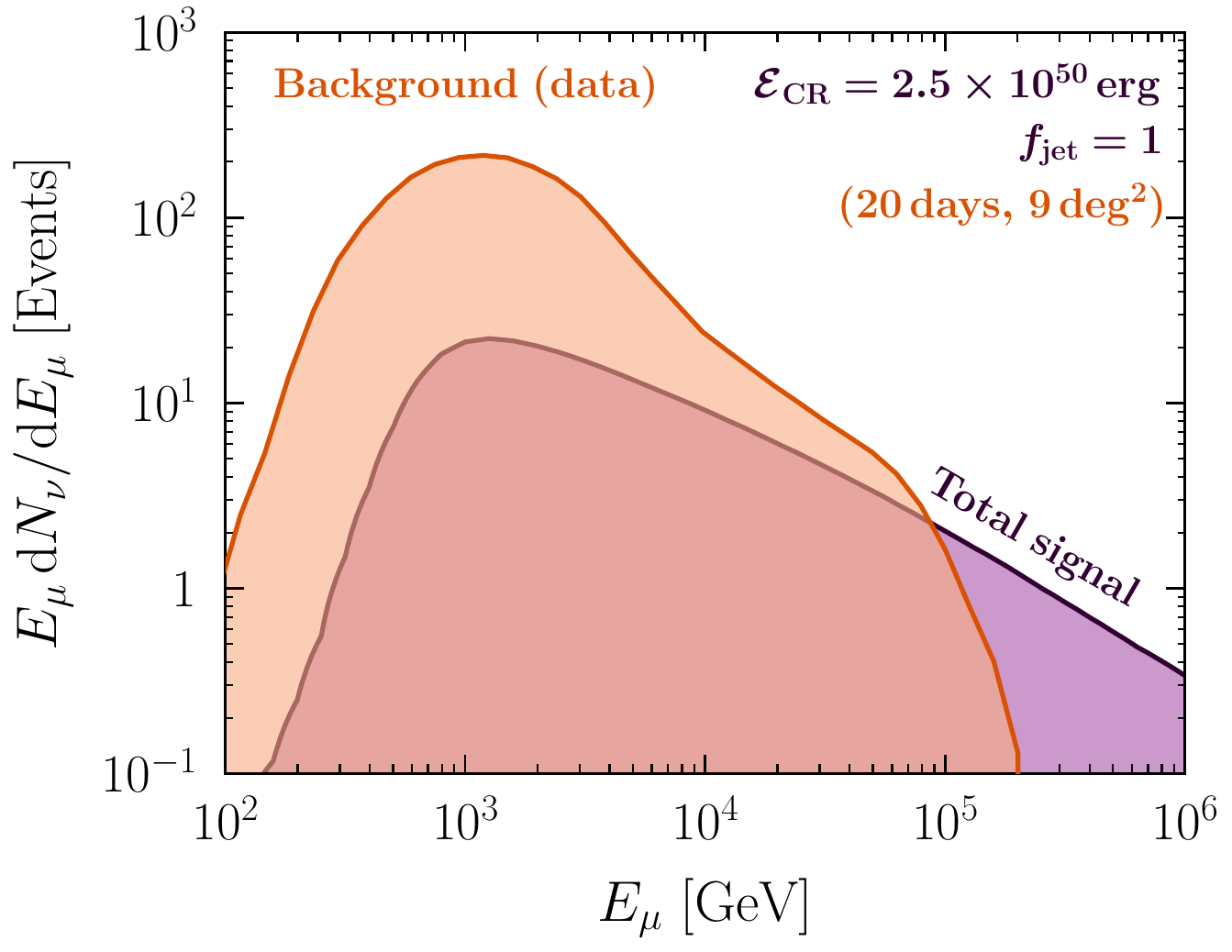}
    \caption{A schematic prediction of the muon-track events produced by our ten-year SNe Ib/c sample as a function of the reconstructed muon energy $E_{\mu}$ in IceCube. Here, the signal events are from the cjSN model with a $E_{\nu}^{-2}$ neutrino spectrum with parameters $\{\mathcal{E}_{\rm CR}, f_{\rm jet}\}$ (detailed in Sec.~\ref{sec_UL}) defined by the upper limit from our analysis. In comparison, we also show the effective number of background events associated with the supernovae [within the effective size of the time (20~days) and spatial window (9 ${\rm deg}^2$); in our analyses, we use more events than shown here (Sec.~\ref{sec_analysis_pdfs})]. {\it The sensitivity of our analysis is mainly driven by the neutrino data with energy above few tens of TeV, where the background becomes negligible compared to the signal.}}
    \label{fig:events_example}
\end{figure}

In this paper, we search for HE neutrinos from cjSNe and study the theoretical implications. Previous studies~\cite{IceCube:2011koo,IceCube:2011wip,Senno:2017vtd,Esmaili:2018wnv,IceCube:2021oiv} have performed searches using early datasets of IceCube and found no association of neutrinos with supernovae. Here we use ten years of IceCube neutrino data~\cite{Abbasi:2021bvk, data_webpage}, in which high-quality events are recorded and have never been analyzed for cjSNe. 
We perform an unbinned maximum-likelihood analysis to search for a statistical correlation between neutrinos and SNe Ib/c, as cjSNe can in principle be observed as SNe Ib/c, where progenitor stars are more massive and typically enclosed by denser extended materials.
Our analyses do not find any excess of neutrinos from SNe Ib/c with respect to the background, from which we set upper limits on cjSN models and their contribution to the total astrophysical neutrino fluxes observed by IceCube. \Cref{fig:events_example} illustrates how muon-neutrino signals from {\it all} SNe Ib/c in our analyzed sample would look in IceCube, assuming the highest cjSN flux allowed by our analysis.

\textit{For the first time}, we take into account physical models of neutrino emission in LP GRBs, instead of simply assuming HE neutrinos to follow power-law spectra. This is important, as the astrophysical neutrino flux may originate from multiple source populations with various neutrino spectra. Moreover, as the current LP GRB sample is highly incomplete and model uncertainties of LP GRBs are largely unconstrained, our survey in model parameters using SNe Ib/c catalogs is meaningful. 
Finally, our conservative limits show that, for most cjSN models we consider, SNe Ib/c can still account for 100\% of the IceCube diffuse neutrino flux. 
Moreover, we find that, even with a very conservative approach, ten years of IceCube data are probing almost all cjSN models, thus implying cjSNe will be robustly tested as HE neutrino sources in the near future.

This paper is organized as follows: In Sec.~\ref{sec_data_model}, we describe the neutrino dataset and the supernova sample we use and introduce the cjSN models we consider. In Sec.~\ref{sec_analysis}, we discuss our likelihood formalism and how we obtain the correlation significance from background simulation. In Sec.~\ref{sec_UL}, we detail the procedures of setting upper limits, including how we simulate neutrino signals from cjSNe and how we look for an excess of signals among background fluctuations. In Sec.~\ref{sec_rslt}, we show our results from single-source and stacking analyses, and we present constraints on cjSN models and their contributions to the HE astrophysical neutrino fluxes observed by IceCube. We further discuss the implications of cjSNe as the origin of HE neutrinos. We then comment on the difference between our results and those in Ref.~\cite{IceCube:2021oiv}. In Sec.~\ref{sec_conclusions}, we conclude our findings with a future roadmap.

\section{Data and Models}
\label{sec_data_model}

\subsection{Ten years of IceCube neutrino data}
\label{sec_data_model_IC}

The IceCube Neutrino Observatory detects neutrinos through the Cherenkov photons emitted by relativistic charged particles produced from neutrino interactions within (starting events) and outside (throughgoing events) the detector~\cite{Achterberg:2006md, ICweb}. The Cherenkov photons trigger the nearby digital optical modules and can form two kinds of basic event topologies: elongated tracks formed by muons, and showers, which look like a round and big blob formed by electrons (electromagnetic shower) or hadrons (hadronic shower). 
The track events, which are dominated by throughgoing tracks, have a much better angular resolution (as good as $< 1^\circ$), though worse energy resolution ($\sim 200\%$ at $\sim 100$~TeV), than the shower events ($\sim$ $10^\circ$--$15^\circ$ and $\sim 15\%$ above 100 TeV)~\cite{ICres}. Thus, track events are suited to searching for point sources.

The data released by the IceCube Collaboration span from April 2008 to July 2018~\cite{Abbasi:2021bvk, data_webpage}. The same data have been used in the ten-year time-integrated neutrino point-source search by the IceCube collaboration~\cite{Aartsen:2019fau}, and in searching for high-energy neutrino emission from radio-bright AGN~\cite{Zhou:2021rhl}. In total, there are 1,134,450 muon-track events. 
The information for each track is provided, including arrival time, angular direction, angular error, and reconstructed energy. The arrival time is given with the precision of $1 \times 10^{-8}$~days ($8.6 \times 10^{-4}$~s).
These ten years of data are grouped into five samples corresponding to different construction phases of IceCube and instrumental response functions, including 1) IC40, 2) IC59, 3) IC79, 4) IC86-I, and 5) IC86-II to IC86-VII. The numbers in the names represent the numbers of strings in the detector on which digital optical modules are deployed.
Distributions of these events in the sky can be found in Figs.~1 and 2 in Ref.~\cite{Zhou:2021rhl}.  
We use the events with declination (Dec) between $-10^\circ$ and $90^\circ$ for the following reasons: First, the events from $\rm Dec < -10^\circ$ (the southern sky with respect to IceCube) have much higher backgrounds from atmospheric muons~\cite{Abbasi:2021bvk}. Second, we find that the given smearing matrices from simulations have statistics that are too low to obtain good enough energy PDFs for our analysis (Sec.~\ref{sec_analysis_pdfs}).

We also process the $19\times2$ double-counted tracks in the dataset found in Ref.~\cite{Zhou:2021xuh} (listed in its Table~III). These events arise from an internal reconstruction error that identifies some single muons crossing the dust layer as two separate muons arriving at the same time and closely matching in direction~\cite{Zhou:2021xuh}. 
This would affect neutrino-source searches, especially transients, as finding two associated events instead of one would be quite different. Thus, we combine the 19 misreconstructed pairs into 19 single events by averaging the directions and summing up the reconstructed energies. We provide the  corrected IceCube neutrino dataset \href{https://github.com/beizhouphys/IceCube_data_2008--2018_double_counting_corrected}{at this URL} \github{beizhouphys/IceCube_data_2008--2018_double_counting_corrected}.

\subsection{Supernova sample}
\label{sec_data_model_SNe}

The supernova sample we use for our analysis is from combining SNe Ib/c from the Open Supernova Catalog~\cite{Guillochon:2016rhj}, the Weizmann Interactive Supernova Data Repository (WISeREP)~\cite{Yaron:2012yc}, and the All-Sky Automated Survey for Supernovae (ASAS-SN)~\cite{Holoien:2016pnj, Holoien:2016hae, Holoien:2017sbn, Holoien:2018kcp}. These catalogs have collected more than 36,000, 20,000, and 1,300 supernovae, respectively, from a variety of astronomical surveys and existing archives.
We further compare our combined supernova sample with the publicly available catalog of bright supernovae~\cite{BSNe, Gal_Yam_2013} and incorporate those that are missed in the above.

Sometimes a supernova is independently discovered by different groups and thus has multiple aliases. This leads to a small fraction of potentially duplicate sources in our sample. To avoid double-counting, we first search for the supernovae with an angular distance smaller than $0.1^{\circ}$. We then merge these supernovae if they are classified as the same type and the differences in their maximal brightness time and redshift are less than 30 days and $10\%$, respectively. The examples of supernova pairs satisfying our criteria are \{SN 2010O, SN 2010P\} and \{SN 2016coi, ASASSN-16fp\}. As these potentially duplicate sources have very similar observational properties, we remove one of them from our sample. 

Finally, we keep the supernovae in our sample only if they were observed at ${\rm Dec} \geq -10^{\circ}$ and have a time window (defined in Sec.~\ref{sec_analysis_pdfs}) overlapping with the uptime of IceCube between April 2008 and July 2018 to match our selected data (Sec.~\ref{sec_data_model_IC}).
In total, our final sample consists of 386 SNe Ib/c, including 30, 36, 36, 36, and 248 for IC40, IC59, IC79, IC86-I, and IC86-II--VII, respectively. We provide the details of our supernova sample \href{https://github.com/SupernovaNus/CZMK2022_SNe_Ibc_sample}{at this URL} \github{SupernovaNus/CZMK2022_SNe_Ibc_sample}.

\begin{figure*}[htpb]
    \centering
    \includegraphics[width=\textwidth]{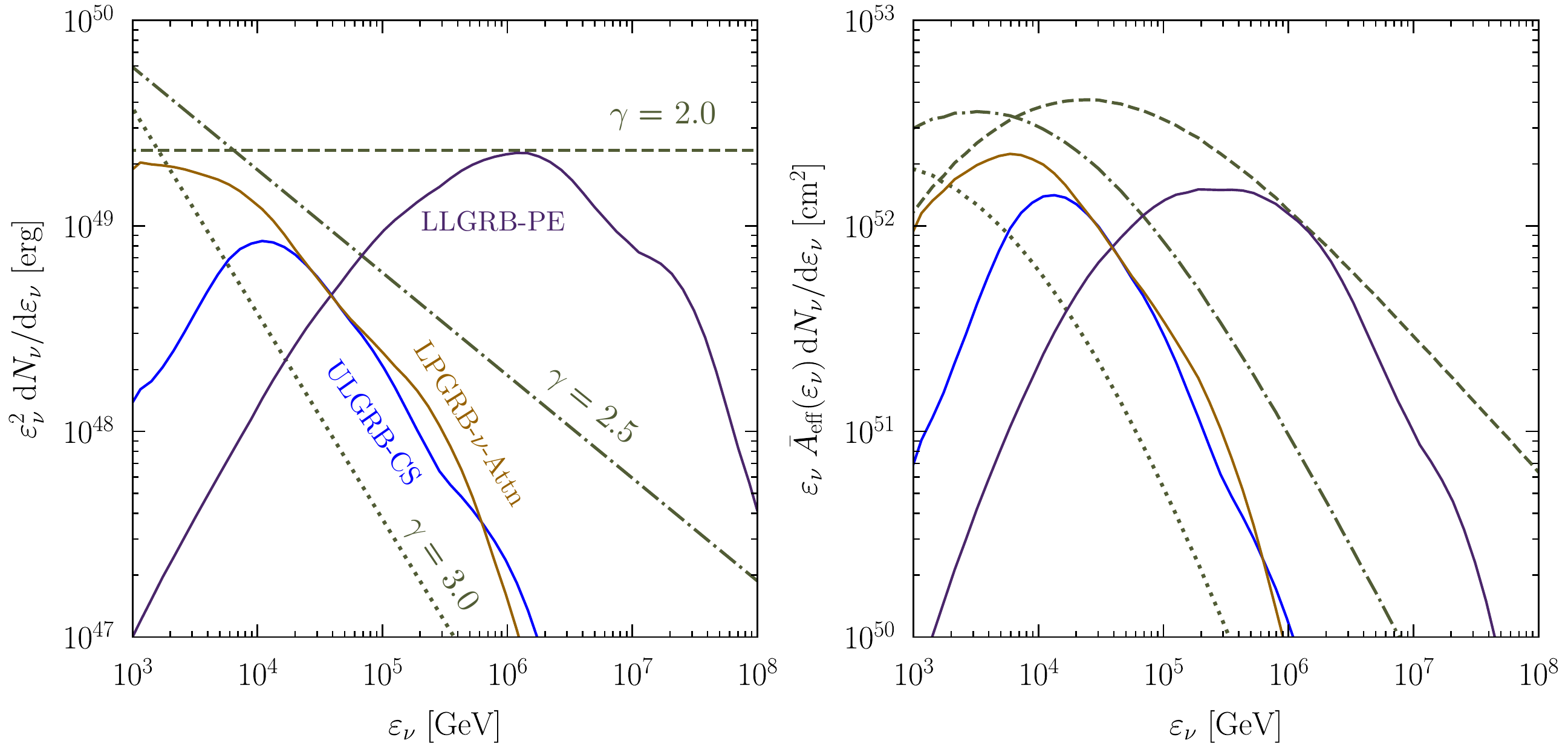}
    \caption{A comparison of single-burst neutrino spectra from the cjSN models we consider. Here, the LLGRB-PE model involves prompt neutrino emission from low-luminosity GRBs~\cite{Murase:2006mm}; the ULGRB-CS model involves neutrinos from cosmic rays accelerated at the collimation shock of ultra-long-duration GRBs~\cite{Murase:2013ffa}; and the LPGRB-$\nu$-Attn model involves neutrinos attenuated in the progenitor star of low-power GRBs~\cite{Carpio:2020app}. All spectra are normalized to have the same isotropic equivalent cosmic-ray energy, $\mathcal{E}_{\rm CR} = 10^{51}$ erg. Both panels use the
    same line style for each model, while in the right panel each spectrum is weighted by the IceCube effective area (averaged over ${\rm Dec} \geq -10^{\circ}$). {\it The realistic modeling of cjSNe takes into account a variety of cooling mechanisms, leading to the suppression in the neutrino spectra and different energy distributions of neutrino events detected by IceCube.}}
    \label{fig:sb_spec}
\end{figure*}

\subsection{cjSN models for neutrino emission}
\label{sec_data_model_nu}

We assume choked jets to be nearly calorimetric sources (except for the suppression factor) so that neutrinos are produced by all available energy $\mathcal{E}_{\rm CR}$ in cosmic rays. The all-flavor neutrino spectrum from a single burst of supernova is thus given by~\cite{Waxman:1998yy, Murase:2019tjj} 
\begin{equation}
    \left.\frac{{\rm d}N_{\nu}(\varepsilon_{\nu};\mathcal{E}_{\rm CR})}{{\rm d}\varepsilon_{\nu}}\right|_{\varepsilon_\nu=0.05\varepsilon_p} \approx \frac{3}{8} f_{\rm sup} \,{\rm min}[1,f_{p\gamma}] \frac{\mathcal{E}_{\rm CR}}{{\mathcal R}_p(\varepsilon_p)} \varepsilon_{\nu}^{-2} \,, \label{eq:sb_nu}
\end{equation}
where the factor $3/8$ is the fraction of energy taken away by neutrinos from charged pions produced in the $p\gamma$ interactions; the energy-dependent suppression factor $f_{\rm sup}(\varepsilon_p)$ accounts for the meson and muon cooling processes that depend on the detailed modeling of choked jets; and the meson production efficiency ${\rm min}[1,\,f_{p\gamma}]$ is set to $1$ for the choked jets as long as the minimal cosmic-ray energy is larger than the pion production threshold. 
As the average fraction of energy transferred from a parent proton to each neutrino after an interaction is $1/20$, we have $\varepsilon_{\nu} \approx (1/20)\varepsilon_p$. ${\mathcal R}_p$ denotes the bolometric correction factor for the cosmic-ray spectrum. The cosmic rays are expected to follow a power-law spectrum~\cite{Waxman:1995dg} (i.e.,~${\rm d}N_p/{\rm d}\varepsilon_p \propto \varepsilon_p^{-\gamma}$), as they are typically accelerated through the first-order Fermi process~\cite{Fermi:1949ee} in the shock. In this case, we have ${\mathcal R}_p(\varepsilon_p) = \ln{(\varepsilon_p^{\rm max}/\varepsilon_p^{\rm min})}$ for $\gamma = 2$ and ${\mathcal R}_p(\varepsilon_p) = {(\gamma-2)^{-1} [1 - (\varepsilon_p^{\rm max}/\varepsilon_p^{\rm min})^{2-\gamma}] (\varepsilon_p/\varepsilon_p^{\rm min})^{\gamma -2 }}$ for $\gamma \neq 2$. When producing our results in Secs.~\ref{sec_rslt_Ecr} and \ref{sec_rslt_diffuse}, we take $(\varepsilon_p^{\rm min}, \varepsilon_p^{\rm max}) = (2 \times 10^3, 2 \times 10^{10})\,{\rm GeV}$ as this leads to efficient neutrino production within $(\varepsilon_{\nu}^{\rm min}, \varepsilon_{\nu}^{\rm max} ) \approx (10^2, 10^9) \, {\rm GeV}$, the energy range that could be detected by IceCube.

We also consider two well-motivated cjSN models \cite{Murase:2006mm,Murase:2013ffa}. Both models assume power-law parent cosmic-ray spectra with $\gamma=2.0$ but the neutrino spectra are different due to various processes of meson and muon cooling in different shocked regions of the jet. In reality,  $f_{p\gamma}$ in \cref{eq:sb_nu} is below unity at low energies, and $f_{\rm sup}<1$ is possible depending on cjSN parameters. 

Our first class of models assumes that the neutrino spectrum follows a power law with the same spectral index ($\gamma$) as the parent proton spectrum. This neglects all complicated mechanisms that could lead to a nontrivial neutrino spectrum. We consider $\gamma = 2.0$, $\gamma = 2.5$, and $\gamma = 3.0$ as three benchmark models, which match the best fits to the ten years of muon-track events~\cite{Stettner:2019tok}, a combination of track and shower events~\cite{IceCube:2015gsk, IceCube:2020acn}, and the 7.5 years of the high-energy starting events (HESEs)~\cite{IceCube:2020wum}, respectively.

Our second class of models takes into account more realistic modeling that links cjSNe with LL GRBs and UL GRBs. We consider three physical models from different detailed considerations of jet propagation inside the progenitor star and energy losses of particles: 1) the LLGRB-PE model (prompt $\nu$ emission from LL GRBs)~\cite{Murase:2006mm}, 2) the ULGRB-CS model (neutrinos from cosmic rays accelerated at the collimation shock of UL GRBs)~\cite{Murase:2013ffa}, and 3) the LPGRB-$\nu$-Attn model (attenuated neutrinos from LP GRBs)~\cite{Carpio:2020app}. Note that the model spectrum used in this work takes into account the inverse-Compton cooling of pions and muons, which slightly affects the flux above $\sim 0.1-1$~PeV compared to the original reference.

The left panel of \cref{fig:sb_spec} shows that for the same $\mathcal{E}_{\rm CR}$, the all-flavor neutrino spectra of a supernova burst from different models can differ by several orders of magnitude at certain energies. The $y$ axis is shown as $\varepsilon_{\nu}^2 {\rm d}N_{\nu}/{\rm d}\varepsilon_{\nu} = 2.3^{-1} \varepsilon_{\nu} {\rm d}N_{\nu}/{\rm d} \log\varepsilon_{\nu}$ so that the area under each curve is proportional to the total energy of neutrinos. 
The right panel of \cref{fig:sb_spec} shows the spectra of detectable muon neutrinos in IceCube, calculated from the curves in the left panel and the average IceCube effective area $\bar{A}_{\rm eff}$ over ${\rm Dec} \geq -10^{\circ}$. For comparison, here we assume that neutrinos are evenly distributed among all flavors.
The $y$ axis is shown as $\varepsilon_{\nu} \bar{A}_{\rm eff} {\rm d}N_{\nu}/{\rm d}\varepsilon_{\nu} = 2.3^{-1} \bar{A}_{\rm eff} {\rm d}N_{\nu}/{\rm d} \log \varepsilon_{\nu}$, so that the area under each curve is proportional to the total number of detected neutrinos. For the same $\mathcal{E}_{\rm CR}$, power-law models (in which there are no cooling effects) or physical models with harder emission spectra between 10\,TeV and 1\,PeV (where IceCube has the best sensitivity) tend to produce more neutrino events in detectors.

\section{Analysis Formalism}
\label{sec_analysis}

\subsection{Likelihood function and test statistic}
\label{sec_analysis_LLH}

To search for a possible correlation between IceCube events and supernovae, we use an unbinned maximum-likelihood method. A similar formalism was commonly used to search for transient neutrino sources by IceCube~\cite{IceCube:2009ror, IceCube:2009xmx, IceCube:2014jkq, IceCube:2016ipa, IceCube:2017amx, IceCube:2017fpg, Fahey:2017zxt, IceCube:2019acm, Kheirandish:2019dii} 
and other experiments~\cite{Super-Kamiokande:2009uwx}. The likelihood function is defined as
\begin{equation}
    \mathcal{L}(s,b) = \prod_{k}^{5} \prod_{j\in k}^{N^k_{\rm SN}} \left[ p_j(s_j,b_j) \prod_{i \in j}^{N_j} \mathcal{L}^k_{ij}(s_j,b_j) \right] \,, \label{eq:llh_total}
\end{equation}
where $k$ labels the five data samples we use in our analysis, $N^k_{\rm SN}$ is the number of supernovae observed during the period of each data sample, 
and $N_j$ is the total number of muon-track events appearing in the temporal and spatial window of the $j$th supernova.
Moreover, $s_j$ and $b_j$ are the signal and background event rates for the $j$th supernova, respectively. In our analysis, $s_j$ is the parameter to be determined by maximizing the likelihood function, and $b_j$ is computed by scaling the number of track events in the supernova's spatial window and outside the temporal window by the ratio between the total times outside and inside its time window.  Details of the windows are given in Sec.~\ref{sec_analysis_pdfs}.

$\mathcal{L}^k_{ij}(s_j,b_j)$ in Eq.~\eqref{eq:llh_total} is the individual likelihood function of a track event $i$ from the $k$th data sample associated with the $j$th supernova,
\begin{equation}
    \mathcal{L}^k_{ij}(s_j,b_j) = \frac{1}{{s_j+b_j}}\left(s_j S^k_{ij} + b_j B^k_{ij} \right)  \,, \label{eq:llh_i}
\end{equation}
with $S_{ij}^k$ being the signal probability density function (PDF) and $B_{ij}^k$ the background PDF; details are given in Sec.~\ref{sec_analysis_pdfs}. 

$p_j(s_j,b_j)$ in Eq.~\eqref{eq:llh_total} is a weighting factor, 
\begin{equation}
    p_j(s_j,b_j) = \frac{(s_j+b_j)^{N_j} {\rm e}^{-(s_j+b_j)} }{N_j!} \, , \label{eq:poisson_j}
\end{equation}
used to take into account the Poisson fluctuations of $N_j$.

The test statistic (TS) is then defined as the ratio of the likelihood function to its value under the null hypothesis
\begin{equation}  
    {\rm TS} \equiv 2 \ln \left[ \frac{\mathcal{L}(s,b)}{\mathcal{L}(0,b)} \right] \,.
\end{equation}
From \cref{eq:llh_total,eq:llh_i,eq:poisson_j} we have
\begin{align}
    {\rm TS} & =  \sum_k^{5} \sum_{j \in k}^{N^k_{\rm SN}} \, 2 \left[ -s_j + \sum^{N_j}_{i \in j} \ln \left( \frac{s_j S^k_{ij}}{b_j B^k_{ij}} + 1 \right)  \right] 
    \label{eq:TS}
    \\
    & \equiv  \sum_k^{5} \sum_{j \in k}^{N^k_{\rm SN}} {\rm TS}^k_j \,,
\end{align}
where ${\rm TS}^k_j$ is the test statistic for a single source covered by the $k$th data sample.
Comparing the TS values from the real with simulated data (details in Sec.~\ref{sec_analysis_simu}), we can get the significance (or $p$-value) of the correlation between the supernova sample and the real data.

\subsection{Signal and background PDFs}
\label{sec_analysis_pdfs}

Both signal and background PDFs for a track event are multiplications of temporal, spatial, and energy PDFs,
\begin{align}
    S^k_{ij} & = S^k_{T,\,ij} \cdot S^k_{{\rm spt},\,ij}\cdot S^k_{E,\,ij}   \,,
    \\
    B^k_{i} & = B^k_{T,\,ij} \cdot B^k_{{\rm spt},\,ij}\cdot B^k_{E,\,ij} \,.
\end{align}
These PDFs describe the expected temporal, spatial, and energy distributions of the events if they are from the sources or the background. For the background PDFs, we calculate them directly from data, as they have enough statistics. 

The signal temporal PDF is taken as a Gaussian distribution in $T_{\rm SN} - T_{\nu}$, the difference between the time of a supernova's maximum brightness and the arrival time of an event,
\begin{align}
    & S^k_{T,\,ij}(T_i|T_{{\rm SN} j})  \nonumber
    \\
    & = \frac{1}{\sqrt{2\pi \sigma_T^2}}\, {\rm exp}\left\{-\frac{[(T_{{\rm SN} j} - T_i) - \lambda_T]^2}{2\sigma_T^2} \right\} \,,
    \label{eq:ST}
\end{align}
where $\sigma_T = 4$ days and $\lambda_T = 13$ days as given below.

In the standard scenarios of cjSNe, high-energy neutrinos and gamma rays are produced simultaneously via the interaction of relativistic jets inside the extended stellar envelope. As it is found that SNe Ib/c correlated with GRBs typically reach their maximal optical brightness after $\simeq 13$ days from the prompt gamma-ray emission~\cite{Cano:2016ccp}, we take $\lambda_T = 13$ days in \cref{eq:ST}. 

When searching for temporal correlation, the time window of each supernova is determined by the $99\%$ confidence interval of \cref{eq:ST}. This gives ${T_{{\rm SN} j} - T_i \in [\lambda_T - 2.58 \sigma_T,\,\lambda_T + 2.58 \sigma_T]}$, so the width of the time window is $\simeq 20$ days. We note that the maximum brightness time itself is not important; instead, whether we can estimate the explosion time with sufficient accuracy is important. For SNe Ib/c, a 20-day time window before the maximum brightness time is a rather conservative choice to include the explosion time for a supernova (see Fig.~10 of Ref.~\cite{Yoshida:2022idr} for uncertainties when all-sky data are available, and Fig.~7 of Ref.~\cite{Barbarino:2020amq} for some example light curves of SNe Ic). This is especially the case for our stacking limits, where the contribution is dominated by the nearby supernovae that are bright enough to estimate the explosion time.
It is almost impossible for the events outside the window to come from the supernova.

For the background temporal PDF, we use a uniform distribution inside the time window and zero outside, 
\begin{equation}
    B^k_{T,\,ij}(T_i|T_{{\rm SN} j}) = \frac{1}{2 \times 2.58  \sigma_T} \,,
    \label{eq:BT}
\end{equation}
because the background event rate from high-energy atmospheric neutrinos and muon interactions are constant. 

The signal spatial PDF is given by the Fisher-Bingham (Kent) distribution~\cite{10.2307/99186,10.2307/2335218,10.2307/2984712},
\begin{equation}
    S^k_{{\rm spt},\,ij}(\mu_{ij}|\sigma_{\nu i}) = \frac{\sigma_{\nu i}^{-2}}{4\pi \,{\rm sinh}\,\left(\sigma_{\nu i}^{-2}\right)} \, {\rm exp}\left( \sigma_{\nu i}^{-2} \mu_{ij} \right) ,\,  
    \label{eq:SS}
\end{equation}
where $\mu_{ij} = \cos(\Delta \psi_{ij})$, with $\Delta \psi_{ij}$ being the angular distance between the track $i$ and supernova $j$, and $\sigma_{\nu i}$ is the reconstructed angular error of a track~\cite{Abbasi:2021bvk, data_webpage}.
\Cref{eq:SS} can be understood as a generalization of the 2D Gaussian distribution with the standard deviation $\sigma_{\nu i}$ on a sphere. For small $\Delta \psi_{ij}$, it reduces to a 2D Gaussian distribution.

Similarly to the time window, we search for spatial correlation through the spatial window defined by the $99\%$ confidence interval of \cref{eq:SS}, which gives $\mu_{ij} \geq \mu_{99}(\sigma_{\nu i})$, where
\begin{equation}
    \mu_{99}(\sigma_{\nu i}) = 1 + \sigma_{\nu i}^2 \ln \left\{ 1 - 0.99 [1 - {\rm exp} (-2 \sigma_{\nu i}^{-2}) ]  \right\} \,.
\end{equation}

The background spatial PDF is a function of Dec, $\delta_i$, as distribution with respect to right ascension (RA) is nearly isotropic, because IceCube is located at the South Pole. We use the same sliding window method as in Ref.~\cite{Zhou:2021rhl} to calculate the background spatial PDF. 
We normalize the background spatial PDF over a Dec from $-10^\circ$ to $90^\circ$, matching the data used in our analysis. The results can be found in Fig.~2 of Ref.~\cite{Zhou:2021rhl}.

The signal energy PDF of an event in sample $k$ can be calculated by
\begin{align}
     & S^k_{E,\,ij} (E^{\rm prx}_i|\delta_i) \nonumber
     \\
     & \propto 
     \int {\rm d}E_{\nu}
     \phi_j(E_{\nu})  
     A^k_{\rm eff}(E_\nu | \delta_i) 
      P^k(E^{\rm prx}_i | E_{\nu},\delta_i)
     \,,
     \label{eq:SE}
\end{align}
where $\phi_j(E_{\nu})$ is the expected muon-neutrino fluence from source $j$ (Sec.~\ref{sec_UL_neutrinos}), 
$A^k_{\rm eff}(E_\nu | \delta_i)$ is the effective area, and $P^k$ is the reconstructed muon energy (energy-proxy, $E^{\rm prx}_i$) distribution for a specific $E_\nu$ and $\delta_i$, given in Refs.~\cite{Abbasi:2021bvk, data_webpage}.

The background energy PDF, $B^k_{E,\,ij} (E^{\rm prx}_i|\delta_i)$, is a normalized distribution of $E^{\rm prx}_i$ as a function of $\delta_i$ and is obtained using a sliding window method with the window size of $\sin \delta_i \pm 0.05$ and $\log_{10}(E^{\rm prx}_i) \pm 0.2$.

\subsection{Significance and simulation}
\label{sec_analysis_simu}

To get the $p$-value (or the statistical significance) for the correlation between the events and supernovae, we need to simulate a large number of synthetic neutrino datasets and calculate the TS.
The $p$-value quantifies the probability that a correlation is observed due to background alone. In our analysis, it is defined as the fraction of the TS of simulated datasets that are larger than the TS of the real data ($\rm TS_{obs}$), and it can be converted to significance under the standard normal distribution. 

The simulation is detailed as follows: We simulate $10^5$ synthetic datasets so that the results converge. For each dataset, it has the same five phases, and for each phase, we simulate the same number of events as the real data. For a simulated event in a specific phase of a specific dataset,
\begin{itemize}
    \item 
    The arrival time and RA are randomly chosen from uniform distributions over the uptime of the phase and $[0,\,2\pi)$, respectively. 
    
    \item The Dec and the energy proxy are randomly chosen according to the background spatial and energy distributions of the real dataset, given in Sec.~\ref{sec_analysis_pdfs}.
    
    \item Given the Dec and energy proxy of a simulated event, the angular error is drawn from the distribution of the angular errors of the events with similar Dec and proxy energy in the real dataset.
    
    \item We keep only the events with $\delta_i \gtrsim -10^{\circ}$, matching our selection for the real data.
\end{itemize}

The signal and background PDFs of the events in the simulated datasets are calculated in the same way as those in the real dataset.

\section{Setting upper limits}
\label{sec_UL}

Here, we detail the procedures for setting upper limits on the parameters of cjSN models. All the models are characterized by the same parameters: the isotropic equivalent cosmic-ray energy, $\mathcal{E}_{\rm CR}$, and the fraction of supernovae that contribute signal neutrinos, $f_{\rm jet}$. The former represents the total energy budget of which cosmic rays can efficiently produce signal neutrinos in the energy range of interest, and the latter accounts for the uncertainties in the fraction of supernovae developing choked jets towards the Earth. The constraints for each model are set from signal-injected simulations, detailed below.

First, we randomly select $f_{\rm jet} N^k_{\rm SN}$ supernovae from our sample and calculate their neutrino spectra for a specific model in Sec.~\ref{sec_data_model_nu}. After convolving with the effective area, we get the detectable neutrino spectra, from which we simulate the neutrino events. 
Second, we simulate each muon-track event from each parent neutrino event using the smearing matrix~\cite{Abbasi:2021bvk, data_webpage},
which takes into account the effects of neutrino interactions, muon energy losses, and detector efficiency. 
We detail the calculation of the number of detectable signal events from individual supernovae in Sec.~\ref{sec_UL_neutrinos}, and the simulations of neutrino events and track events in Sec.~\ref{sec_UL_tracks}.
In every signal-injected simulation, which corresponds to a specific $\{\mathcal{E}_{\rm CR}, f_{\rm jet}\}$ and specific model, the above procedure is done for each supernova in our sample and we inject the simulated signal track events into a randomly selected simulated neutrino dataset (Sec.~\ref{sec_analysis_simu}).
Next, we calculate the TS value for each signal-injected simulation following the procedure in the previous section and compare them with the TS-value distribution of the simulated datasets (Sec.~\ref{sec_analysis_simu}). 

\begin{figure}[t]
    \centering
    \includegraphics[width=\columnwidth]{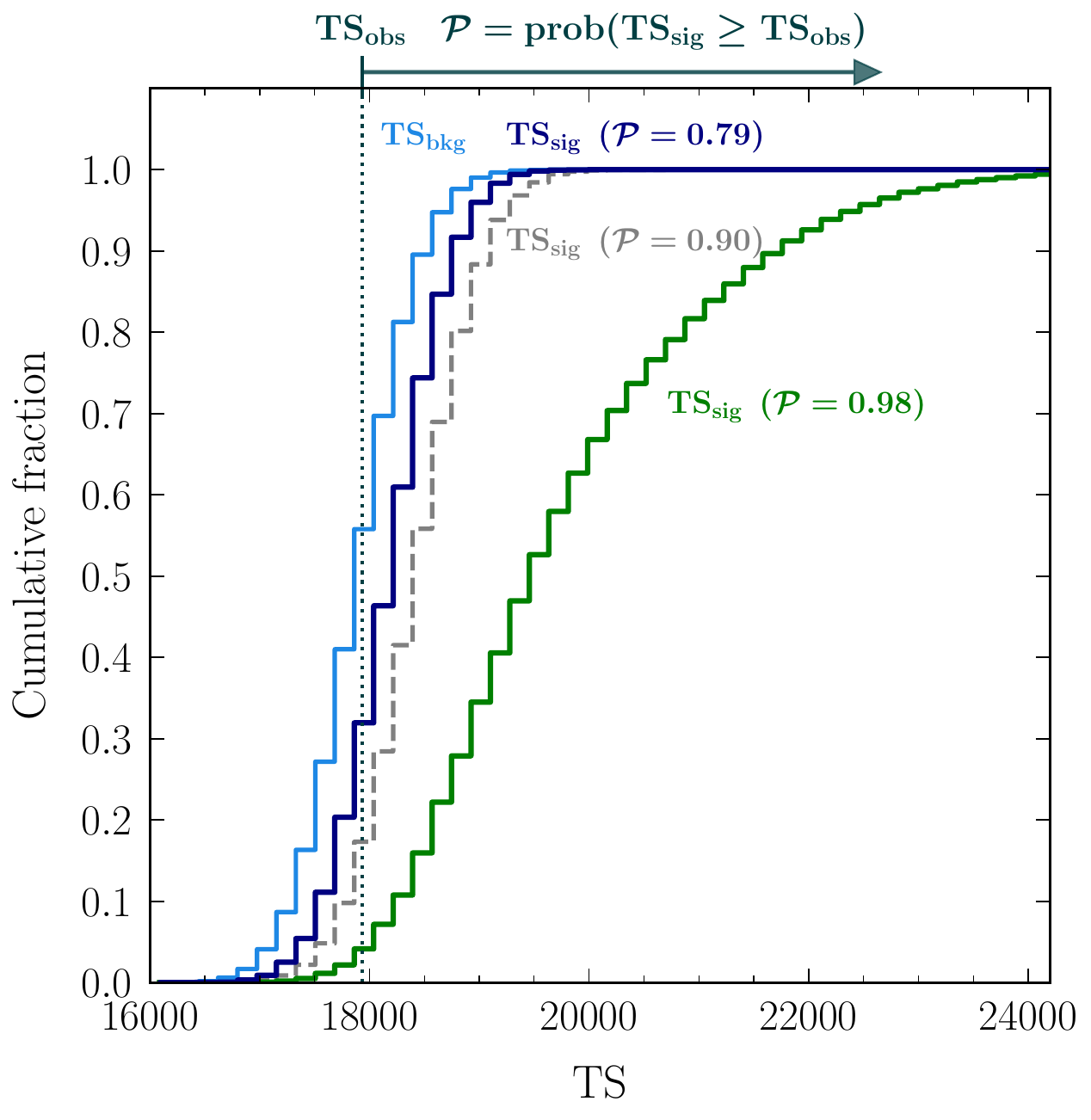}
    \caption{Cumulative TS-value distributions of the data with (labeled with $\rm TS_{sig}$) and without ($\rm TS_{bkg}$) injected signals. 
    Here we use our power-law model with $\gamma = 2.0$ as an example. The exclusion confidence level of a signal hypothesis is defined by the probability $\mathcal{P}$ of finding ${\rm TS_{sig} \geq TS_{obs}}$, and our exclusion boundary is set by $\mathcal{P} = 0.90$ (the grey dashed curve).
    {\it We exclude signal hypotheses in which cjSN models lead to an excess of significance, i.e., $\mathcal{P}>90\%$}. For example, the green curve is excluded while the dark blue curve is not.}
    \label{fig:TS_dist}
\end{figure}

The above procedure is repeated for different $\{\mathcal{E}_{\rm CR}, f_{\rm jet}\}$ values and models. For each model, to produce our limits, we scan over $\{\mathcal{E}_{\rm CR}, f_{\rm jet}\}$ parameter space, in which the TS-value distribution of each point is obtained by performing $10^4$ simulations. We have checked that our results converge well.

\Cref{fig:TS_dist} exemplifies the exclusion for our power-law model with index $\gamma=2$.
Given \{$\mathcal{E}_{\rm CR}$, $f_{\rm jet}$\},
we follow IceCube's approach~\cite{Aartsen:2014aqy,Aartsen:2016qcr,Aartsen:2017wea,IceCube:2017fpg,IceCube:2022rlk} in which the exclusion confidence level for a cjSN model is determined by the probability of finding ${\rm TS}$ greater than ${\rm TS_{obs}}$. For example, the three signal hypotheses (labeled ${\rm TS_{sig}}$) we show in the figure, obtained with $\{\mathcal{E}_{\rm CR},f_{\rm jet}\} = \{1.6 \times 10^{50} \,{\rm erg}, 1.0\}$, $\{4.0 \times 10^{50} \,{\rm erg}, 0.6\}$, and $\{6.3 \times 10^{51} \,{\rm erg}, 0.1\}$, are excluded at $79\%$, $90\%$, and $98\%$ confidence level, respectively.

\subsection{Neutrino fluence from cjSN}
\label{sec_UL_neutrinos}

The sensitivity of our analysis to a model strongly depends on its total signal-neutrino flux. For the scenarios of cjSNe, this is closely related to $\mathcal{E}_{\rm CR}$, as the neutrinos are mainly produced through the charged pion decay following the meson production from the $p \gamma$ interactions of accelerated protons in the jet~\cite{Waxman:1998yy}. For the $k$th data sample, we calculate the number of detected signal neutrino events by the sum of contributions from the individual supernovae, which is
\begin{equation}
    N^k_{\nu} = \sum^{f_{\rm jet} N^k_{\rm SN}}_j \int_{E_{\nu}^{\rm min}}^{E_{\nu}^{\rm max}} {\rm d}E_{\nu} A^k_{\rm eff}(E_{\nu} | \delta_{{\rm SN}j}) \phi_j(E_{\nu};\mathcal{E}_{\rm CR})\,,
    \label{eq:N_nu_inj}
\end{equation}
where, with the flux equally distributed in flavors, the muon-neutrino fluence from a single supernova burst at luminosity distance $d_j$ (redshift $z_j$) is related to \cref{eq:sb_nu} by
\begin{equation}
    \phi_j(E_{\nu};\mathcal{E}_{\rm CR}) = \frac{(1+z_j)}{4\pi d_j^2} \cdot \frac{1}{3} \frac{{\rm d}N_{\nu}(\varepsilon_{\nu};\mathcal{E}_{\rm CR})}{{\rm d}\varepsilon_{\nu}}\,,
\end{equation}
the observed neutrino energy is $E_{\nu} = (1+z_j)^{-1}\varepsilon_{\nu}$, and the Dec of the signal neutrinos is the same as that of their parent supernova, $\delta_{{\rm SN}j}$. $E^{\rm min}_{\nu}$ and $E^{\rm max}_{\nu}$ are determined by the energy range of the accelerated protons at the source.

\begin{figure*}[t!]
    \includegraphics[width = 0.75\textwidth]{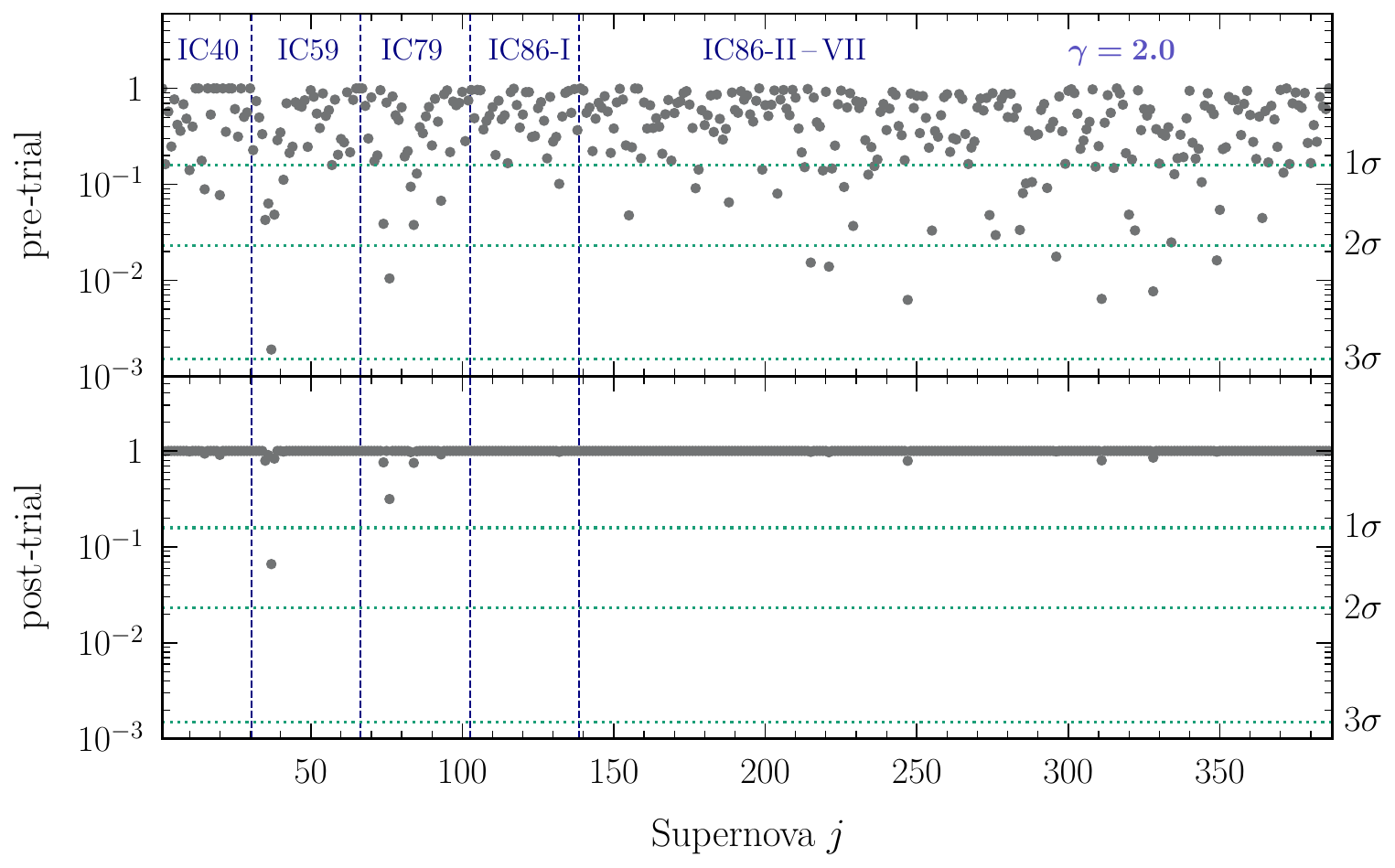}
    \caption{Pretrial and posttrial $p$-values of the 386 SNe Ib/c in our sample, given a power-law model with $\gamma = 2.0$. The corresponding significance is shown over the $y$ axis on the right. {\it No significant sources are observed}.
    }
    \label{fig:p_values}
\end{figure*}

\subsection{Simulate detected track events}
\label{sec_UL_tracks}

The detected track events from the neutrino events are simulated with the following procedure:
\begin{itemize}
    \item The arrival time $T_i$ is randomly chosen from the signal temporal PDF in \cref{eq:ST}.
    
    \item The Dec of the neutrino event, $\delta_i$, is set to be that of the supernova, $\delta_{{\rm SN}j}$.
    
    \item The energy of a detected signal neutrino arriving at Earth, $E^{\rm det}_{\nu i}$, is randomly drawn from the energy spectrum of detected neutrinos; that is, $A^k_{\rm eff}(E_{\nu} | \delta_i) \phi_j(E_{\nu})$ in \cref{eq:N_nu_inj}.
    
    \item Given the $E^{\rm det}_{\nu i}$ and $\delta_i$ of a neutrino event, the proxy energy $E^{\rm prx}_i$, the angular error, and the point spread function (PSF, the angle between the parent neutrino's direction and the reconstructed muon direction) of its daughter track event are randomly drawn by following the smearing matrix.
    
    \item Given the PSF, the angular distance, RA, and Dec of the track are drawn by following the signal spatial PDF with $\sigma_{\nu} = {\rm PSF}$ in \cref{eq:SS}.
    
\end{itemize}

\section{Results and Discussion}
\label{sec_rslt}

In this section, we present our results. In Sec.~\ref{sec_rslt_analysis}, we show our single-source and stacking analysis results. Then, we show the upper limits on the choked-jet model parameters in Sec.~\ref{sec_rslt_Ecr} and their contributions to the diffuse high-energy neutrino flux in Sec.~\ref{sec_rslt_diffuse}.

\subsection{Single-source and stacking analysis results}
\label{sec_rslt_analysis}

In our single-source analysis, the pretrial $p$-value for individual supernovae ($p^k_{{\rm pre},\,j}$) is determined by the probability of finding ${\rm TS}^k_{{\rm bkg},\,j} \geq {\rm TS}^k_{{\rm obs},\,j}$ in the test statistic distribution under background-only hypothesis. However, a low pretrial $p$-value does not necessarily mean a discovery but could be caused by the statistical fluctuations of the background. This can be taken into account by the posttrial $p$-value, which can be calculated by
\begin{equation}
    p^k_{{\rm post},\,j} = 1 - \left( 1 - p^k_{{\rm pre},\,j} \right)^{N^k_{\rm SN}} \,.
    \label{eq:p_post_j}
\end{equation}

Figure~\ref{fig:p_values} shows the pretrial (upper panel) and posttrial (lower panel) $p$-values of all the SNe Ib/c in our sample, along with the corresponding significance in the unit of standard normal deviations. 
The results show that none of the supernovae in our sample have significant neutrino emission.
The highest pretrial significance is $\simeq 2.9\sigma$ (SN 2009hy), but its posttrial significance is only $\simeq 1.5\sigma$.
Here, we only report the $p$-values based on the power-law spectrum with $\gamma = 2.0$, as we find that changing the neutrino spectrum only marginally affects the significance of each source. 
We provide the $p$-values for each supernova given all cjSN models \href{https://github.com/SupernovaNus/CZMK2022_SNe_Ibc_sample}{at this URL} \github{SupernovaNus/CZMK2022_SNe_Ibc_sample}.

Table~\ref{tab:p_value} shows our stacking analysis results. The posttrial $p$-values are calculated by
$  p_{\rm post} = 1 - \left( 1 - p_{\rm pre} \right)^{6}$, where 6 is the total number of spectra/models we test.
The results show that for all the models we consider, the data are consistent with the background-only hypothesis.

\begin{table}[htbp]
\caption{Pretrial and posttrial $p$-values and significance from our stacking analysis for different cjSN models. 
}
\label{tab:p_value}
\medskip
\renewcommand{\arraystretch}{1.1} 
\centering
\begin{tabular*}{\columnwidth}{ccc}
\hline\hline
Model &  $p_{\rm pre}$ (significance) 	& \makebox[0.2cm]{}	$p_{\rm post}$ (significance)	
\\
\hline\hline
$\gamma = 2.0$  &  0.583 (0)   &  0.995 (0)
\\
$\gamma = 2.5$  &   0.517 (0)  &  0.987 (0)
\\
$\gamma = 3.0$  &   0.450 (0.1$\sigma$)  &  0.972 (0)
\\
\hline
LLGRB-PE~\cite{Murase:2006mm}  &  0.641 (0)  &  0.998 (0)
\\
ULGRB-CS~\cite{Murase:2013ffa}  &  0.613 (0)  &   0.997 (0)
\\
LPGRB-$\nu$-Attn~\cite{Carpio:2020app} & 0.556 (0)  & 0.992 (0)
\\
\hline\hline
\end{tabular*}
\end{table}

\subsection{Stacking constraints on cjSN model parameters}
\label{sec_rslt_Ecr}

\begin{figure}[t]
    \centering
    \includegraphics[width=\columnwidth]{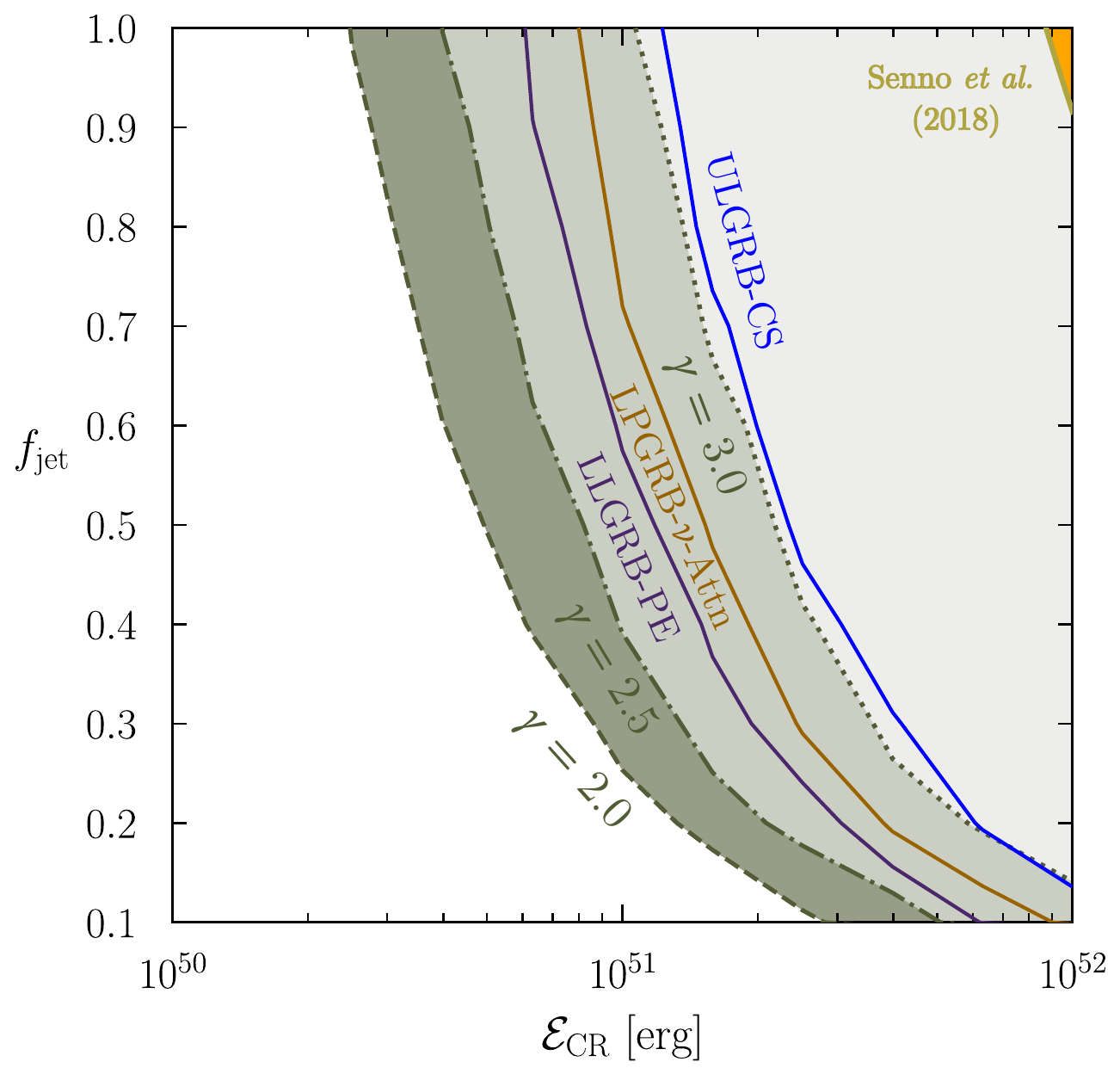}
    \caption{Upper limits on $\mathcal{E}_{\rm CR}$ and $f_{\rm jet}$ for each cjSN model. Also shown in the upper-right corner is the existing 90\% confidence-level limit of $\gamma = 2.0$ obtained by previous work~\cite{Senno:2017vtd} using one year of track events. {\it Realistic models are generally allowed to have larger total energy in cosmic rays, as they are less efficient in producing detectable neutrinos.}}
    \label{fig:fj_Ecr}
\end{figure}

Figure~\ref{fig:fj_Ecr} shows our upper limits on $\mathcal{E}_{\rm CR}$ and $f_{\rm jet}$ at a 90\% confidence level for different cjSN models. The parameter space above the curves is excluded.
Among all the scenarios we consider, the power-law model with $\gamma = 2.0$ has the strongest constraints, because it produces the most signal events in the energy range to which IceCube detectors are most sensitive. 
For the same reason, power-law models with softer spectra are less constrained, and the limit on the ULGRB-CS model is relatively weak, as it produces the smallest amount of neutrinos in that energy range. 

Our limit on $\mathcal{E}_{\rm CR}$ for the power-law model with $\gamma = 2.0$ is stronger than the existing one-year limit in Ref.~\cite{Senno:2017vtd} by more than an order of magnitude at $f_{\rm jet} = 1$.
The improvement is expected, because our supernovae sample covered by the ten-year IceCube dataset is more than ten times larger than that used in Ref.~\cite{Senno:2017vtd}.

Theoretically, $\mathcal{E}_{\rm CR}$ and to $f_{\rm jet}$ are completely degenerate. In that case, the contours should all follow $\sim 1/\mathcal{E}_{\rm CR}$. However, there is a small deviation at small $f_{\rm jet}$ values. This is due to the limited number of supernovae in our sample, and for small $f_{\rm jet}$ the uncertainty of our limits on $\mathcal{E}_{\rm CR}$ is larger. 
Also, it should be noted that our limits are mainly driven by nearby supernovae. Thus, the degeneracy holds when $\sum_k f_{\rm jet} (N^k_{\rm SN} \cdot f_{\rm nb}) > 1$, where $f_{\rm nb}$ is the fraction of nearby sources in our samples. During our signal-inject procedure, we find that more than $50\%$ of the signal events are contributed by the supernovae with luminosity distance $d_j\lesssim 16\,{\rm Mpc}$, which account for $\lesssim 2\%$ of all the supernovae in our sample, so we set $f_{\rm nb} = 0.02$. With $\sum_k N^k_{\rm SN} = 386$, we expect that the limits are obtained for  $f_{\rm jet} \gtrsim 0.1$. 
This implies that with the present supernova sample, the limits on cjSNe models are rather meaningful only if the jets have relatively wide opening angles, $\theta_j\gtrsim0.3$ rad. To test cjSN models with smaller opening angles---e.g., $\theta_j\sim0.1$ rad---better samples with more nearby supernovae are required.  

\subsection{Upper limits on diffuse neutrino flux}
\label{sec_rslt_diffuse}

\begin{figure*}[htbp]
    \centering
    \includegraphics[width=0.75\textwidth]{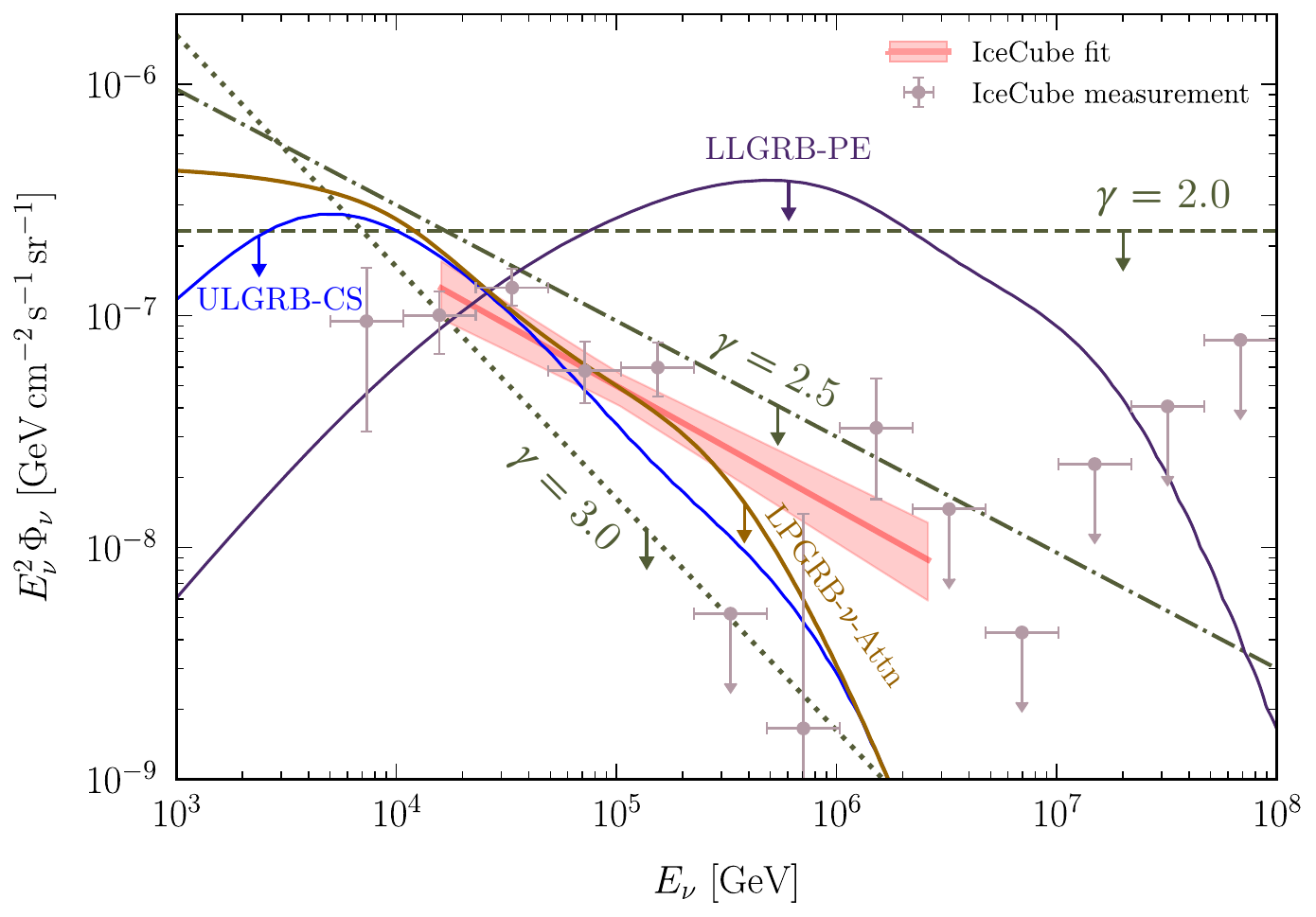}
    \caption{Upper limits on the all-flavor diffuse neutrino flux contributed from each of our cjSN models. The red band corresponds to the IceCube measurement of the all-sky astrophysical neutrino flux and its uncertainties, assuming an unbroken single power law: $E_{\nu}^2\Phi_{\nu} = \Phi_0 (E_{\nu}/ 100 \,{\rm TeV})^{2-\gamma}$, with $\Phi_0 = \left(4.98^{+0.75}_{-0.81} \right) \times 10^{-8} \, {\rm GeV\,cm^{-2}\,s^{-1}\,sr^{-1}}$ and $\gamma = 2.53 \pm 0.07 $~\cite{IceCube:2015gsk,IceCube:2020acn}. The data points with error bars and the red band represent the best-fit results from the 2010--2015 shower data of IceCube~\cite{IceCube:2020acn}. Note that the diffuse neutrino flux measurement obtained from the 9.5-year muon track data of IceCube~\cite{IceCube:2021uhz} leads to the same conclusion. {\it Because of the high local rate density of supernovae, most cjSN models we consider remain viable to explain the diffuse flux of astrophysical neutrinos measured by IceCube.}}
    \label{fig:diffuse_limit}
\end{figure*}

Now we convert our limits on $\mathcal{E}_{\rm CR}$ and $f_{\rm jet}$ to upper limits on the contribution to diffuse neutrino flux from all the cjSN models. The diffuse flux can be calculated by (e.g., \cite{Murase:2015xka})
\begin{align}
    \Phi_{\nu}(E_{\nu}) & =  \frac{c}{4\pi H_0} \int_0^{z_{\rm max}} {\rm d}z \int_{L_{\rm min}}^{L_{\rm max}} {\rm d}L_{\gamma}  \hfill 
    \label{eq:diff_flux_1}
    \\
    & \times  \frac{{\rm d} \rho(z)/{\rm d}L_{\gamma}}{\sqrt{ \Omega_{\Lambda} + \Omega_{\rm m} (1+z)^3}} \left.\frac{{\rm d}N_{\nu}(\varepsilon_{\nu};\mathcal{E}_{\rm CR})}{{\rm d}\varepsilon_{\nu}}\right|_{\varepsilon_{\nu}=0.05\varepsilon_p} \,, \nonumber
\end{align}
where $\varepsilon_{\nu} \equiv (1+z)E_{\nu}$, ${\rm d}\rho(z)/{\rm d} L_{\gamma}$ is the luminosity function that takes into account the distribution of on-axis choked-jet supernova rate density $\rho(z)=f_{\rm jet}R(z)$ with respect to the observed GRB luminosity, $H_0$ is the Hubble constant, and $\{\Omega_{\Lambda}, \Omega_{\rm m}\}$ are the cosmological energy density parameters. Combining the above equation with \cref{eq:sb_nu} gives the all-flavor diffuse neutrino flux by~\cite{Bahcall:1999yr,Murase:2015xka}
\begin{align}
    E_{\nu}^2 & \Phi_{\nu}(E_{\nu}) \sim  7.6 \times 10^{-8} \,{\rm GeV\,cm^{-2}\,s^{-1} \, sr^{-1}}  \nonumber 
    \\
    &  \times  f_{\rm sup}(\varepsilon_p) \, {\rm min}[1,f_{p\gamma}] \frac{\xi_z}{3} \left.\left[\frac{{\mathcal R}_p(\varepsilon_p)}{10}\right]^{-1} \right|_{\varepsilon_p \sim 20\varepsilon_{\nu} } \nonumber
    \\
    & \times  \left( \frac{f_{\rm jet} \,\mathcal{E}_{\rm CR} \cdot R_{\rm L}}{10^{51}\,{\rm erg} \cdot 10^3\,{\rm Gpc^{-3}\,yr^{-1}}} \right)   
    \,, \label{eq:diff_flux_2}
\end{align}
where $\xi_z$ and $R_{\rm L}\equiv R(0)$ are the redshift evolution factor~\cite{Waxman:1998yy, Bahcall:1999yr} and the true local rate density (i.e., volumetric rate) of the supernovae that harbor choked jets (including off-axis ones), respectively. Equation~\eqref{eq:diff_flux_2} shows that an upper limit on $f_{\rm jet} \mathcal{E}_{\rm CR}$ directly constrains the maximal neutrino flux we can detect. 

\Cref{fig:diffuse_limit} shows the upper limits (at a $90\%$ confidence level) on the all-flavor diffuse neutrino flux from SNe Ib/c for different cjSN models. The bounds are set via the $\mathcal{E}_{\rm CR}$ limits for $f_{\rm jet} = 1$ from \cref{fig:fj_Ecr}. (Note that our limits are sensible for $f_{\rm jet}\gtrsim0.1$, as discussed below.) 
To be conservative and consistent with our supernova sample, we take $R_{\rm L} = R_{\rm Ib/c} \sim  2 \times 10^4 \,{\rm Gpc}^{-3}\,{\rm yr}^{-1}$, the local rate density of SNe Ib/c~\cite{Li:2010kc,Melinder:2012nv}. 
We find that the power-law models with $\gamma = 2.0$ and $\gamma = 2.5$, ULGRB-CS, LLGRB-PE, and LLGRB-$\nu$-Attn can still explain all the diffuse astrophysical neutrino fluxes observed by IceCube. While the soft power-law model with $\gamma = 3.0$ cannot contribute more than $59\%$ of the observed diffuse neutrino flux.
Although we focus on cjSN models, SNe Ib/c may be aided by newborn pulsar winds, which could be emitters of neutrinos in the PeV--EeV range~\cite{Murase:2009pg, Fang:2014qva, Fang:2018hjp}.  Our analysis results can be applied to such models but the limits are still above the model predictions. 

It is worth noting that, although we constrain physical cjSN models motivated by various rarer types of supernovae or GRBs, one should not simply downscale our diffuse flux limits by applying the local rate density of rarer source classes to Eq.~\eqref{eq:diff_flux_2}. This is because our SN Ib/c sample does not necessarily include any GRBs or hypernovae. If hypernovae or GRBs (including LL GRBs) only compose a small fraction of the entire supernova sample (i.e., $R_{\rm L}<R_{\rm Ib/c}$), our approach described in Sec.~\eqref{sec_rslt_Ecr} may not give a sensible limit. This is the case even if all SNe Ib/c harbor jets (i.e., $R_{\rm L}=R_{\rm Ib/c}$), because only a fraction of them have jets pointing to us. In the current analysis with $\sim400$ supernovae, we can apply the limits down to $f_{\rm jet} \sim 0.1$, and more supernova samples are necessary to probe cjSN models with smaller jet opening angles. 
This point is important in order not to misinterpret the results of the stacking analyses. The choice of the source rate density in the stacking analysis presented here must always be consistent with the sample of sources one takes into account. 

\subsection{Comparison with prior work}
\label{sec_comparison}

Reference~\cite{IceCube:2021oiv} performed a similar stacking analysis with cjSNe and seven years of track events. The result shows that a power-law neutrino spectrum with $\gamma = 2.5$ and a similar width of time window contributes no more than $16.4\%$ of the diffuse astrophysical neutrino flux. The constraint is about 10 times stronger than our ten-year limit for $\gamma = 2.5$ in \cref{fig:diffuse_limit}, as Ref.~\cite{IceCube:2021oiv} constrained the isotropic equivalent energy to be $\mathcal{E}_{\rm CR} < 8 \times ({\rm limit~on~total~muon~neutrino~energy}) \sim 4 \times  10^{49} \,{\rm erg}$, which is also $\sim 10$ times smaller than our result in \cref{sec_rslt_Ecr} ($\gamma = 2.5$, $f_{\rm jet} = 1$: $\mathcal{E}_{\rm CR} < 4.0 \times 10^{50} \,{\rm erg}$). 

A few reasons may cause such discrepancy. 
First, Ref.~\cite{IceCube:2021oiv} analyzed a subsample of 19 nearby supernovae that accounts for 70\% the total neutrino flux. We find that, for all models, our limits on $\mathcal{E}_{\rm CR}$ in Fig.~\ref{fig:fj_Ecr} (with $f_{\rm jet} = 1$) and the diffuse neutrino flux in Fig.~\ref{fig:diffuse_limit} could be stronger by a factor of $3$--$4$ if we analyze the subsample that dominates the total neutrino flux, because the total background would be reduced.
We emphasize that, for our main results, we do not select any subsamples in the analyses, so our limits are regarded as robust and ``conservative'' constraints on cjSN models. Importantly, even if all supernovae harbor jets, only a fraction ($f_{\rm jet}$) of them would point to us. In reality, the viewing angle of choked jets can be small, so that $f_{\rm jet} \ll 1$. We caution that the stacking analysis with a small subsample will greatly decrease the sensitivity for $f_{\rm jet} \ll 1$. If none of the on-axis events are included in the nearby sample, sensible constraints cannot be placed, as the viewing angle of choked jets is unknown. 

Second, there is an important difference between the source catalog of Ref.~\cite{IceCube:2021oiv} and ours: their catalog includes some nearby Type IIb supernovae (e.g.,~SN 2011dh), while we only include Type Ib/c supernovae as cjSNe are typically expected to be detected as these types. We find that the large flux produced by the additional SNe IIb would affect the final constraints by a factor of $\sim 2$.

Third, Ref.~\cite{IceCube:2021oiv} constructed signal energy PDFs through the smearing matrices from IceCube's internal Monte Carlo simulations, which are different from the publicly released smearing matrices we use, and they may also lead to an improvement in the final sensitivity. 

Finally, the likelihood formalism used in Ref.~\cite{IceCube:2021oiv} is slightly different from ours. We use the same formalism as IceCube's search for neutrino emission from transient sources like GRBs~\cite{IceCube:2009ror, IceCube:2009xmx, IceCube:2014jkq, IceCube:2016ipa, IceCube:2017amx} and fast radio bursts~\cite{IceCube:2017fpg, Fahey:2017zxt, Fahey:2017zxt, IceCube:2019acm, Kheirandish:2019dii}, which includes a Poisson weighting factor [Eq.~\eqref{eq:poisson_j}]. Reference~\cite{IceCube:2021oiv} used a different weighting method, as some of their analyses have much longer time windows (up to 1000~days).

Furthermore, we note that the constraints with steep spectral indices such as $\gamma=2.5$ lead to aggressive limits on the contribution to the diffuse neutrino flux. As shown in Fig.~\ref{fig:diffuse_limit}, the limits for harder indices are weaker for neutrinos at energies above $\sim30$~TeV. Also, models with steep spectra down to TeV energies cannot make a significant contribution to the diffuse flux in light of energetics of GRBs and SNe, and most of the viable models in the literature have a low-energy break or cutoff. Reference~\cite{IceCube:2021oiv} did not provide the constraints on models with harder indices or more realistic neutrino spectra, and their limits are used for constraining models with much caution.

All in all, our results are complementary to Ref.~\cite{IceCube:2021oiv} in various aspects. We focus on setting robust constraints on various physical neutrino emission models of cjSNe with $f_{\rm jet} \leq 1$ using SNe Ib/c, while Ref.~\cite{IceCube:2021oiv} puts constraints on power-law models with $f_{\rm jet}=1$ using both SNe Ib/c and SNe IIb, but we stress that these limits should be interpreted with much caution from the theoretical perspectives.

\section{Conclusions}
\label{sec_conclusions}

The detection of TeV--PeV astrophysical neutrino flux by IceCube is a breakthrough in neutrino physics, astrophysics, and multimessenger astronomy. These neutrinos are unique probes of neutrino physics at high energies in the standard model~\cite{Glashow:1960zz, Seckel:1997kk, Alikhanov:2015kla, IceCube:2017roe, Bustamante:2017xuy, Zhou:2019vxt, Zhou:2019frk, IceCube:2020rnc, IceCube:2021rpz, Zhou:2021xuh, Xie:2023qbn, Plestid:2024bva} and beyond~\cite{Lipari:2001ds, Cornet:2001gy, Beacom:2002vi, Beacom:2006tt, Yuksel:2007ac, Murase:2012xs, Feldstein:2013kka, Esmaili:2013gha, Murase:2015gea,Shoemaker:2015qul, Bustamante:2016ciw, IceCube:2016dgk, Coloma:2017ppo, Denton:2018aml, IceCube:2018tkk, Arguelles:2019ouk, Esteban:2021tub}. 
They also directly probe the interiors of astrophysical dense environments opaque to photons and will cast light on the long-standing problems of the origin of HE cosmic rays and their acceleration mechanisms. To fully exploit the exclusive opportunities brought by HE astrophysical neutrinos, searching for and studying their sources are the essential steps. 

Previous studies indicated that most astrophysical neutrino sources should be optically thick to gamma rays \cite{Murase:2015xka}. Among promising gamma-ray dark candidates, core-collapse supernovae with jets choked by surrounding materials are especially interesting~\cite{Murase:2013ffa,Tamborra:2015fzv,Senno:2015tsn}, as they 1) naturally explain the potential connections between SNe Ib/c; hypernovae; and classical, LL, and UL GRBs, and 2) are efficient in producing HE astrophysical neutrinos. 

In this paper, we studied whether cjSNe can be the dominant source contributing to the diffuse astrophysical neutrinos detected by IceCube. We used the unbinned maximum-likelihood method to search for the associations between IceCube neutrinos and SNe Ib/c. Compared to existing searches, our analysis takes advantage of ten years of IceCube neutrino data. We also collected SNe Ib/c from publicly available supernova catalogs covering the same period. Importantly, for the first time, we looked for the neutrino signals of physical cjSN models in IceCube data. These models take into account a variety of time-dependent cooling processes for cosmic rays and mesons in LP GRBs, which are well motivated but have never been constrained by data. 

For all the cjSN models we considered, our single-source analysis (Fig.~\ref{fig:p_values}) and stacking analysis (Table~\ref{tab:p_value}) found no significant correlation between the muon-track events and any SNe Ib/c in our source sample with respect to the backgrounds. Thus, we put upper limits on, $\mathcal{E}_{\rm CR}$, the total energy budget of cosmic rays, and $f_{\rm jet}$, the fraction of SNe Ib/c that has choked jets pointing to us for different cjSN models. 
Our ten-year limit improves the prior one-year limit in Ref.~\cite{Senno:2017vtd} by more than an order of magnitude (Fig.~\ref{fig:fj_Ecr}).
Moreover, the limits of both parameters are generally sensitive to the efficiency of a cjSN model in producing detectable neutrinos by IceCube (see the right panel of \cref{fig:sb_spec}), which can vary by a factor of up to $\sim 5$ among the models we consider. 

We set the corresponding upper limits of the cumulative astrophysical neutrino fluxes contributed by all SNe Ib/c for each model (Fig.~\ref{fig:diffuse_limit}). In contrast to strongly constrained transient neutrino sources such as classical GRBs~\cite{IceCube:2009ror, IceCube:2009xmx, IceCube:2014jkq, IceCube:2016ipa, IceCube:2017amx}, jetted TDEs \cite{Stein:2019ivm}, gamma-ray blazars~\cite{Aartsen:2016lir, Hooper:2018wyk, Yuan:2019ucv, Luo:2020dxa, Smith:2020oac}, and radio-loud AGN~\cite{Plavin:2020emb, Plavin:2020mkf,Zhou:2021rhl}, 
SNe Ib/c have a much higher volumetric rate (see Table 1 of Ref.~\cite{Murase:2019tjj}), so they could potentially make a large contribution to HE neutrino fluxes whether they have choked jets or not. 

Future prospects of testing cjSNe as neutrino sources are very promising. From \cref{fig:diffuse_limit}, we clearly see that most of our conservative limits obtained from ten years of data are close to the measured astrophysical neutrino flux in IceCube. Given that IceCube-Gen2~\cite{IceCube-Gen2:2020qha} will be at least 5 times more sensitive than IceCube, IceCube-Gen2 will effectively strengthen our stacking limits and critically constrain most cjSN models with just two years of operation. Additionally, there will be much more high-quality data from other observatories such as KM3NeT~\cite{Adrian-Martinez:2016fdl}, Baikal-GVD~\cite{Baikal-GVD:2018isr}, P-ONE~\cite{P-ONE:2020ljt}, and TRIDENT~\cite{Ye:2022vbk}, which are sensitive to different parts of the sky. Importantly, the Vera C. Rubin Observatory’s upcoming Legacy Survey of Space and Time (LSST)~\cite{LSSTScience:2009jmu, LSST:2008ijt} will increase the detection rate of supernovae by more than an order of magnitude ($> 10^5$ core-collapse supernovae per year~\cite{Lien:2009db, Lien:2010yb}). The improvement of the redshift completeness of nearby supernova samples at low redshift ($z \lesssim 10^{-2}$) will dramatically boost the sensitivities of future stacking analyses; furthermore, an increase of the observed supernovae at higher redshift will be beneficial to constraining the cjSN models with $f_{\rm jet} \ll 0.1$. This is crucial because the typical beaming factor of GRB jets is believed to be $f_{\rm jet}\sim0.01$. The realtime optical follow-up programs~\cite{Pan-STARRS:2019szg,ASAS-SN:2022gst,Stein:2022rvc} will also be sensitive to the counterparts of choked-jet supernovae~\cite{IceCube:2016cqr}. In addition, multimessenger coincident searches combined with ultraviolet, x-ray and gamma-ray observations~\cite{AyalaSolares:2019iiy,Reichherzer:2021pfe} will also be useful for LL GRBs that may lead to shock breakout emission.
Together, core-collapse supernovae with choked jets have reached the fork of being discovered or ruled out as dominant neutrino sources. Our insights into the interrelationship between the most elusive high-energy particles and the stellar explosions within dense environments will drastically grow in the near future.

\section*{Acknowledgments}
We thank John Beacom, Jose Carpio, Ali Kheirandish, Michael Larson, William Luszczak, Jannis Necker, and the anonymous referee for helpful discussions. 
P.-W.~C. was supported by NSF Grant No.\ PHY-2012955 to John Beacom and the Studying Abroad Fellowship of Ministry of Education, Taiwan. B.~Z. and M.~K. were supported by NSF Grant No.\ 2112699 and the Simons Foundation. 
K.~M. was supported by NSF Grants No.~AST-1908689, No.~AST-2108466, and No.~AST-2108467, and KAKENHI Grants No.~20H01901 and No.~20H05852.
The computational resources of this work were provided by the Ohio Supercomputer Center~\cite{OhioSupercomputerCenter1987}.

\bibliography{references}

\end{document}